\documentclass[twocolumn]{aastex631}

\submitjournal{AJ}

\shorttitle{Stellar kinematics in W5}
\shortauthors{Lim et al.}

\begin{document}

\title{The kinematics of young stellar population in the W5 region of the 
Cassiopeia OB6 association: implication on the formation process of stellar 
associations}

\correspondingauthor{Beomdu Lim}
\email{blim@kongju.ac.kr}

\author[0000-0001-5797-9828]{Beomdu Lim}
\affiliation{Department of Earth Science Education, Kongju National University, 
56 Gongjudaehak-ro, Gongju-si, Chungcheongnam-do 32588, Republic of Korea}
\affiliation{Korea Astronomy and Space Science Institute, 776 
Daedeok-daero, Yuseong-gu, Daejeon 34055, Republic of Korea}
\affiliation{Earth Environment Research Center, Kongju National University, 56 Gongjudaehak-ro, Gongju-si, Chungcheongnam-do 32588, Republic
of Korea}
\author[0000-0002-5097-8707]{Jongsuk Hong}
\affiliation{Korea Astronomy and Space Science Institute, 
776 Daedeok-daero, Yuseong-gu, Daejeon 34055, Republic of Korea}
\author[0000-0001-6993-1506]{Jinhee Lee}
\affiliation{Korea Astronomy and Space Science Institute, 
776 Daedeok-daero, Yuseong-gu, Daejeon 34055, Republic of Korea}
\author[0000-0001-6842-1555]{Hyeong-Sik Yun}
\affiliation{Korea Astronomy and Space Science Institute, 
776 Daedeok-daero, Yuseong-gu, Daejeon 34055, Republic of Korea}
\author[0000-0002-2013-1273]{Narae Hwang}
\affiliation{Korea Astronomy and Space Science Institute, 776 Daedeok-daero, Yuseong-gu, Daejeon 34055, Republic of Korea}
\author[0000-0002-6982-7722]{Byeong-Gon Park}
\affiliation{Korea Astronomy and Space Science Institute, 776 Daedeok-daero, Yuseong-gu, Daejeon 34055, Republic of Korea}



\begin{abstract} 
The star-forming region W5 is a major part of the Cassiopeia 
OB6 association. Its internal structure and kinematics may 
provide hints of the star formation process in this region. 
Here, we present a kinematic study of young stars in W5 using 
the Gaia data and our radial velocity data. A total 490 out 
of 2,000 young stars are confirmed as members. Their spatial 
distribution shows that W5 is highly substructured. We identify 
a total of eight groups using the k-means clustering algorithm. 
There are three dense groups in the cavities of H {\scriptsize II} 
bubbles, and the other five sparse groups are distributed at the 
ridge of the bubbles. The three dense groups have 
almost the same ages (5 Myr) and show a pattern of expansion. 
The scale of their expansion is not large enough to account for the 
overall structure of W5. The three northern groups are, in fact, 
3 Myr younger than the dense groups, which indicates the independent 
star formation events. Only one group of them shows the signature of 
feedback-driven star formation as its members move away from the 
eastern dense group. The other two groups might have formed in a 
spontaneous way. On the other hand, the properties of two southern 
groups are not understood as those of a coeval population. Their 
origins can be explained by dynamical ejection of stars and multiple star 
formation. Our results suggest that the substructures 
in W5 formed through multiple star-forming events in a giant molecular cloud.
\end{abstract}



\section{Introduction} \label{sec:1}
Star formation takes place on a few parsecs to several hundreds of 
parsecs scales in a hierarchical way \citep{EEPZ00}. Stellar associations 
are the superb laboratories to study star formation process on 
such different spatial scales as they are the prime star-forming sites 
distributed along the spiral arm structure in the host galaxies \citep{BEM96,LL03,G18}. 
OB associations are particularly interesting 
stellar systems because they contain a number of massive stars 
\citep{A47}, which are rare in the solar neighborhood. OB associations 
are, in general, composed of a single or multiple stellar clusters 
and a distributed stellar population \citep{B64,KAG08,LNGR19,LHY20}.  
This internal structure may be closely associated with their formation 
processes.

Expansion of stellar clusters has been steadily detected in 
many associations \citep{KHS19,LNGR19,LHY20,LNH21}. 
These findings seem to be the key features to understand 
the unboundedness of associations according to a classical 
model for the dynamical evolution of embedded clusters 
after rapid gas expulsion \citep{T78,H80,LMD84,KAH01,
BK13,BK15}. Based on the observational data, \citet{LHY20} 
suggested that the young stellar population distributed 
over 20 pc in the W4 region of the Cassiopeia OB6 association 
originates from escaping stars from the central open cluster 
IC 1805. 

However, cluster expansion alone cannot explain the origin 
of substructures commonly found in stellar associations. Such 
substructures are composed of stellar groups (or subclusters) 
\citep{KFG14} that are kinematically distinct \citep{LNGR19,LNH21,
LNH22}. The formation of substructures can naturally be 
explained by star formation along filaments in almost all 
turbulent clouds \citep{A15}. A range of gas densities leads 
to different levels of star formation efficiencies. High-density 
regions are the sites of cluster formation \citep{BSC11,K12}. 
Gas clumps have different sizes and velocity dispersions 
depending on their virial states, which is observed as the 
so-called size-line width relation \citep{L81}. There are 
attempts to detect this signature from substructures in 
stellar associations \citep{LNGR19,WKR20}.

Since \citet{EL77} proposed the so-called collect and 
collapse scenario, a number of observational studies 
have reported the signatures of feedback-driven star 
formation, such as the morphological relationship 
between remaining gas structures and young stellar objects (YSOs), 
and their age sequences \citep[etc.]{FHS02,
SHB04,ZPD07,KAG08,LSK14b}. Recently, the physical 
causality between the first and the second generations of 
stars was assessed by using gas and stellar kinematics 
\citep{LSB18,LNH21}. Meanwhile, a series of theoretical 
work showed that feedback from massive stars predominantly 
suppresses subsequent star formation by dispersing 
remaining clouds \citep{DEB12,DEB13,DHB15}. This result is 
supported by recent observations \citep{YLL18,YLK21}. The 
cores in the $\lambda$ Orionis cloud exposed to a massive 
O-type star have higher temperatures, lower densities, lower 
masses, smaller sizes, and lower detection rates of dense gas tracers 
(N$_2$H$^+$, HCO$^+$, and H$^{13}$CO$^+$) than those in 
the adjacent star-forming clouds Orion A and B, which implies 
the former cloud has less favorable conditions for core formation 
than the others. Therefore, further observational studies 
are required to test the collect and collapse scenario.

The massive star-forming region (SFR) W5, which is a major 
part of the Cassiopeia OB6 association, is an ideal target to 
study the formation process of stellar associations. The 
previously determined distances to this SFR range from 
1.7 kpc to 2.3 kpc \citep{S55,JHI61,BF71,GG76,M72,LGM01,
CPO11,LSK14a}. Its age is younger than 5 Myr \citep{KM03,
KA11,LSK14a}. This SFR is divided into the two regions W5 
East and W5 West as it is surrounded by two giant H 
{\scriptsize II} bubbles \citep{KM03}. The major sources 
of ionization are four O-type stars, BD +60 586 
(O7.5V), HD 17505 (O6.5III((f))), HD 17520 (O9V), and 
HD 237019 (O8V) \citep{MCW55,CL74,HGB06}. The presence 
of numerous YSOs have also been 
confirmed using extensive imaging surveys \citep{CHS00,KAG08}. 
Most YSOs form clusters, while some are spread over 
several tens of parsecs \citep{KAG08} as seen in many 
associations \citep{B64,KL14}. 

Early studies of the bright-rimmed cloud IC 1848A (W5A/S201) 
at the border of the giant H {\scriptsize II} region suggested 
that star formation in the cloud had been triggered by the 
expansion of the H {\scriptsize II} region \citep{LW78,
TTH80}. \citet{WHL84} also found another possible site of 
feedback-driven star formation at the northern cloud (W5NW). 
The double-peaked $^{13}$CO ($J = 1 - 0$) line they 
observed was interpreted as a result of the passage of a shock 
driven by the ionization front. Later, it was found that a 
number of YSOs and cometary nebulae were distributed 
along the H {\scriptsize II} bubble \citep{KM03,KAG08,
KAK08}. In addition, YSOs far away from the ionizing 
sources tend to be at an earlier evolutionary stage of 
protostars \citep{KAG08}. These results were interpreted 
by feedback-driven star formation models \citep{EL77,SWK82}.

The presence of multiple clusters, distributed stellar 
population, and the young stars distributed along the 
border of H {\scriptsize II} regions suggest that this 
SFR might have been formed through multiple 
processes. The absence of kinematic information 
has hindered our understanding of its formation process. 
However, the parallax and proper motion (PM) data obtained 
from the Gaia mission \citep{gaia16} along with radial 
velocities (RVs) allow us to evaluate the membership 
of young stars and further investigate their kinematic 
properties. In this study, we aim to understand the 
formation process of this SFR. Data that we used are 
described in Section~\ref{sec:2}. In Section~\ref{sec:3}, the 
scheme of genuine members is addressed. We present 
the results of this study in Section~\ref{sec:4} 
and discuss the star formation process within W5 
in Section~\ref{sec:5}. Finally, our results are 
summarized in Section~\ref{sec:6} along with our conclusions.

\section{Data} \label{sec:2}
\subsection{Selection of member candidates}\label{ssec:21}
Most OB associations are distributed along the Galactic plane 
\citep{W20}, and therefore a large number of field interlopers 
are observed together in the same field of view. The member 
selection is a procedure of crucial importance to obtain reliable 
results as emphasized by our previous observational studies (e.g., 
\citealt{LHY20,LNH21,LNH22}). We selected members through 
two steps. First, the candidates of young star members were 
identified using several spectrophotometric criteria. 
Second, the final member candidates can be selected 
in the parallax and PM cuts.

We first gathered four different catalogues. 
Massive O- and B-type stars found in SFRs are probable 
member candidates because of their short lifetime, especially 
for O-type stars. We obtained the lists of such O- and B-type 
stars from several databases of MK classification \citep{WOE00,
R03,S09,MSM13}. A catalogue of 192 O- and B-type stars 
in W5 region was created after some duplicates were removed. 
\citet{KAK08} published a list of 17,771 infrared sources 
distributed over W5 region. We took only 2,062 
sources showing infrared excess. Later, a catalogue of 408 YSO 
candidates was released by \citet{KA11}. This catalogue contains 
the spectral types and $H\alpha$ equivalent widths of 
the stars. We considered stars with $H\alpha$ equivalent widths 
smaller than $-10$\AA \ and $0$\AA \ as $H\alpha$ emission 
stars and candidates, respectively. The last catalogue 
contains a total of 567 members in W5 West selected 
using $UBVI$ and $H\alpha$ photometry \citep{LSK14a}.

We cross-matched the four catalogues to create a master 
catalogue of member candidates. All O- and B-type stars 
were found in the catalogue of \citet{KAK08} except one. 
A total 564 out of 567 member candidates from \citet{LSK14a} 
have the infrared counterparts. Among the three candidates 
without infrared counterparts, two candidates are H$\alpha$ 
emission stars, and the other one is an early-type star. 
Since they are highly probable members, we added these four 
sources to the master catalogue. All the YSO candidates from \citet{KA11} were included in the infrared 
source list of \citet{KAK08}. The master catalogue contains 
a total of 2376 member candidates, of which 2000 have counterparts in 
the catalogue of Gaia Early Data Release 3 (EDR3; 
\citealt{gedr3}).

\begin{figure}[t]
\epsscale{1.0}
\plotone{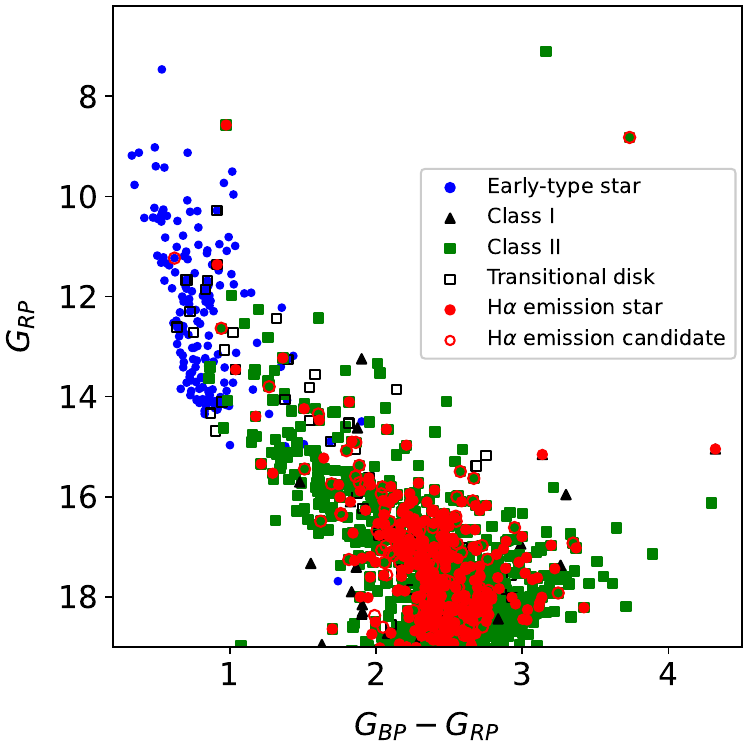}
\caption{Color-magnitude diagram of stars in W5 region. Blue dots, 
black triangles, green squares, black open squares, red dots, and red 
open circles represent early-type stars, Class I, Class II, YSOs 
with a transitional disk, H$\alpha$ emission stars, and H$\alpha$ 
emission star candidates, respectively. The photometric data of 
these stars were taken from Gaia EDR3 \citep{gedr3}.}\label{fig1}
\end{figure}

The parallaxes of Gaia EDR3 have zero-point offsets as 
a function of magnitude, color, and ecliptic latitude 
\citep{LBB21}. We corrected such offsets for the parallaxes 
of individual member candidates using the public Python 
code (\citealt{LBB21}; \url{https://gitlab.com/iccub/public/gaiadr3_zeropoint}). 
In the catalogue of member candidates, we did not use 
stars with negative parallaxes or close companion 
(duplication flag = 1 or RUWE $>$ 1.4) and stars without 
astrometric parameters in analysis. Figure~\ref{fig1} 
displays the color-magnitude diagram (CMD) of the member candidates. 

\subsection{Radial velocities}\label{ssec:22}
We performed multi-object spectroscopic observations of 273 YSO 
candidates on September 3, October 30 in 2020, and October 22, 27, and 29 in 2021 
using high-resolution ($R \sim 34,000$) multi-object spectrograph 
Hectochelle \citep{SFC11} on the 6.5m telescope of the MMT 
observatory. All the spectra were taken with the RV31 filter that 
covers a spectral range of 5150 to 5300 \AA \ in a $2\times2$ 
binning mode. For one observation setup, several tens of fibers 
were assigned to the YSO candidates, and the others were directed 
toward blank sky to obtain sky spectra. The exposure time for 
each frame was set to 35 minutes. A minimum of three frames were 
taken for the same observation setup to eliminate cosmic rays 
and achieve as high a signal-to-noise ratio as possible. For 
calibration, dome flat and ThAr lamp spectra were also obtained 
just before and after the target observation. 

We reduced the raw mosaic frames using the IRAF\footnote{Image
Reduction and Analysis Facility is developed and distributed by the
National Optical Astronomy Observatories, which is operated by the
Association of Universities for Research in Astronomy under operative
agreement with the National Science Foundation.}/{\tt MSCRED}
packages following standard reduction procedures. One-dimensional 
spectra were subsequently extracted from the reduced frames using 
the {\tt dofiber} task in the IRAF/{\tt SPECRED} package. Target
spectra were then flattened using dome flat spectra. The solutions
for the wavelength calibration obtained from ThAr spectra
were applied to both target and sky spectra. 

Some spectra were affected by the scattered light because our observations 
were conducted under bright sky condition. The scattered light 
was unevenly illuminated over the field of view ($1^{\circ}$ in 
diameter), resulting in a spatial variation of sky levels. Hence, 
we constructed a map of sky levels for a given setup following the 
procedure used in our previous study (see \citealt{LNH21} for detail). 
Target spectra were subtracted by sky spectra scaled at given target 
positions. The sky-subtracted spectra for the same target were then 
combined into a single spectrum. Finally, all target spectra were 
normalized by using continuum levels traced from a cubic spline 
interpolation. We rejected the spectra of 115 targets from 
subsequent analysis. Among them, the spectra of 108 targets have signals 
close to the sky background levels, and therefore the signals of 
these spectra were insufficient to measure RVs. The spectra of 
six targets were dominated by continuum, and that of the other 
one is dominated by emission lines.

We measured the RVs of the rest 158 YSO candidates 
using a cross-correlation technique. Synthetic stellar spectra 
for the solar abundance and $\log g = 4$ were generated in a wide 
temperature range of 3,500 to 10,000 K using {\tt SPECTRUM 
v2.76} \citep{GC94}\footnote{http://www.appstate.edu/~grayro/spectrum/spectrum.html} 
based on a grid of the ODFNEW model atmospheres \citep{CK04}. 
These synthetic spectra were used as template spectra. We derived 
the cross-correlation functions between the synthetic spectra and 
the observed spectra of the YSO candidates with {\tt xcsao} task 
in the \textsc{RVSAO} package \citep{KM98}. The velocities at the 
strongest correlation peaks were adopted as the RVs of given 
YSO candidates. The errors on RVs were estimated using the equation 
as below \citep{KM98}:
\begin{equation}
\epsilon(\mathrm{RV}) = {3w \over 8(1+h/\sqrt{2}\sigma_a)}
\end{equation}
\noindent where $w$, $h$, and $\sigma_a$ represent the full widths 
at half-maximum of cross-correlation functions, their amplitudes, 
and the root mean square of antisymmetric components, respectively. 
Rapidly rotating stars, in general, have large uncertainties 
in RVs because they have large full widths at half-maximum of 
cross-correlation functions. Also, the RV errors exponentially 
increase as the r-statistics of cross-correlation functions ($r = h/\sqrt{2}\sigma_a$; \citealt{TD79}) decrease. Indeed, it was confirmed that the cross-correlation functions with r-statistics smaller than 6 yield 
very large errors of RVs ($> 5$ km s$^{-1}$). We thus excluded some 
RV measurements where the r-statistics of cross-correlation 
functions is less than 6. The median error of the RVs is about 
1.2 km s$^{-1}$. The RVs of YSO 
candidates were then converted to velocities in the local 
standard of rest frame using the \textsc{IRAF}/{\tt RVCORRECT} task. 

\begin{figure}[t]
\epsscale{1.0}
\plotone{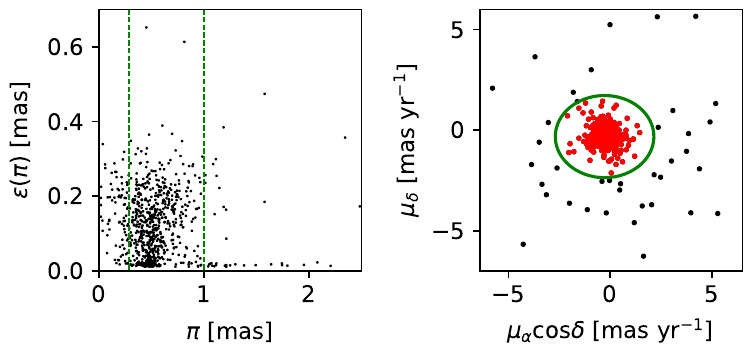}
\caption{Parallax (left) and PM (right) distributions of 
member candidates. The left panel displays the parallaxes and 
their associates errors. We plot stars that are brighter than 
18 mag in $G_{\mathrm{RP}}$ band and have parallaxes greater 
than three times the associated errors. In the left panel, 
dashed lines indicate the boundary used to search for genuine 
members between 1.0 and 3.5 kpc. The right panel exhibits the PM distributions 
of member candidates between the 1.0 and 3.5 kpc. 
Only stars with parallaxes greater than their 
associated errors are considered for analysis. The ellipse in the 
right panel shows the region confined within five times the 
standard deviation from the weighted mean PMs, where the 
inverse of the squared PM error is used as the weight value. 
The selected members are shown by red dots.}\label{fig2}
\end{figure}

\begin{figure}[t]
\epsscale{1.0}
\plotone{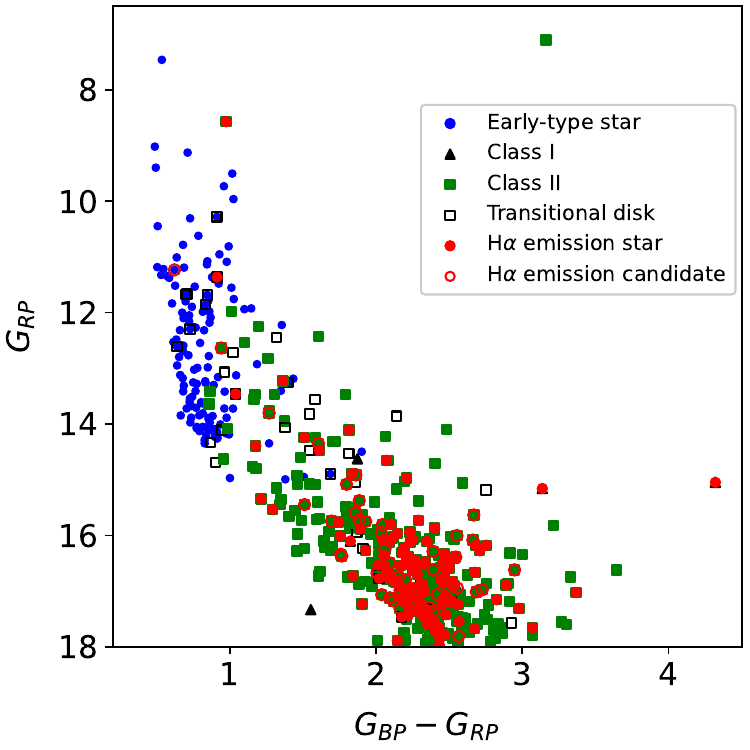}
\caption{CMD of the selected members. The symbols are the 
same as in the Figure~\ref{fig1}.}\label{fig3}
\end{figure}

\section{Member selection} \label{sec:3}
We selected the member candidates based on the spectrophotometric 
properties of young stars. However, there may be a number 
of nonmembers in the catalogue of member candidates (see \citealt{LHY20,
LNH21,LNH22}). For instance, the two bright infrared sources 
($G_{\mathrm{RP}} < 10$ mag and $G_{\mathrm{BP}} - G_{\mathrm{RP}} > 3$) 
in Figure~\ref{fig1} are probably asymptotic giant branch stars 
in the Galactic disk because they are too bright to be the pre-main-sequence 
members of this SFR. It is, thus, necessary to filter out 
additional nonmembers using the Gaia parallaxes and PMs of stars. 

In order to exclude stars with very large measurement errors in 
parallax and PM, we used stars that are brighter 
than 18 mag in $G_{\mathrm{RP}}$ band and have parallaxes 
greater than three times the associated errors. Note that a 
total of 863 candidates are fainter than 18 mag. The left panel 
of Figure~\ref{fig2} displays the parallax distribution of the member 
candidates. Most member candidates have parallaxes smaller than 
1 mas ($d > 1$ kpc). The distances determined from previous studies 
range from 1.7 to 2.3 kpc \citep{S55,JHI61,BF71,GG76,M72,LGM01,
CPO11,LSK14a}. However, we considered candidates between 1.0 kpc and 
3.5 kpc to contain as many probable members as possible. Their 
PMs distribution is shown in the right panel of Figure~\ref{fig2}.

The PM distribution shows the strong concentration of 
member candidates around ($\mu_{\alpha}\cos\delta$, 
$\mu_{\delta}$) = (0 mas yr$^{-1}$, 0 mas yr$^{-1}$). 
Most of them may be genuine members. In order to remove 
PM outliers, the statistical clipping method described in 
\citet{LNH22} was applied for the member candidates.

\begin{figure}[t]
\epsscale{1.0}
\plotone{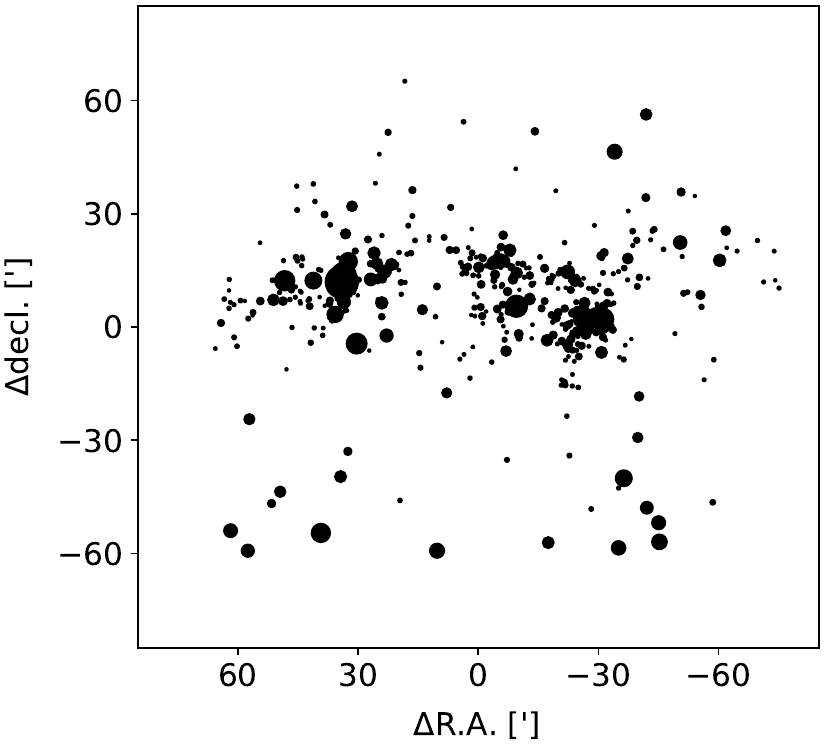}
\caption{Spatial distribution of members in W5. The 
size of dots is proportional to the brightness of 
individual stars. The positions
of stars are relative to the reference coordinate
R.A. = $02^{\mathrm{h}} \  54^{\mathrm{m}} \ 45\fs80$,
decl. = $+60^{\circ} \ 22^{\prime}  \ 04\farcs3$ (J2000).}\label{fig4}
\end{figure}

\begin{figure}
\epsscale{1.0}
\plottwo{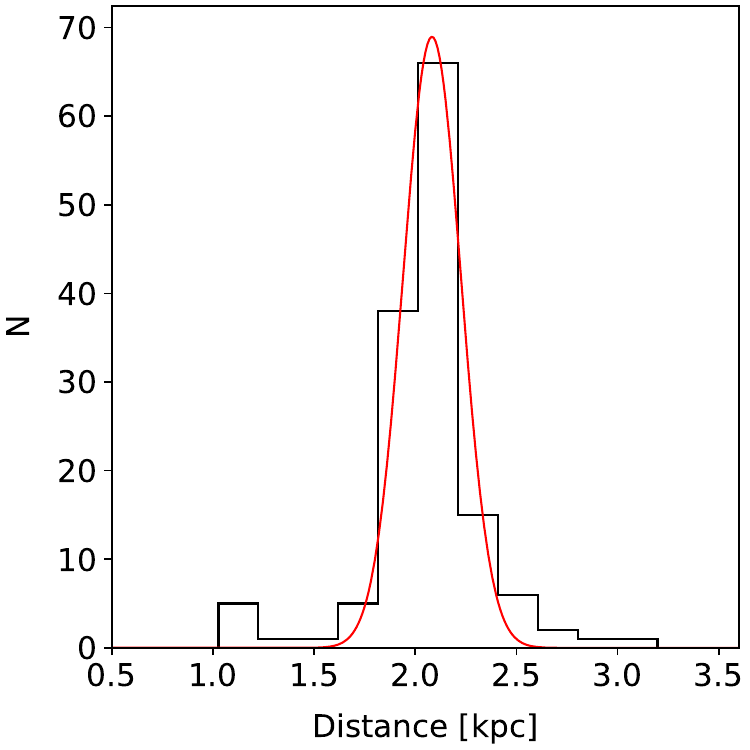}{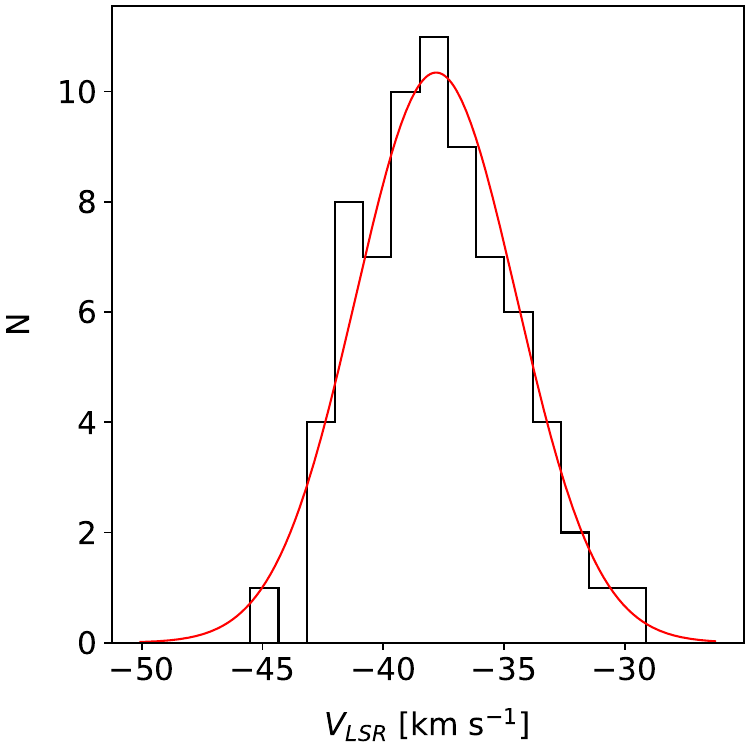}
\caption{Distribution of distances (left) and RVs (right). In order 
to compute the reliable distance to W5, we used members with 
parallaxes larger than 10 times the associated errors. The bin 
sizes of 0.2 kpc and 1.2 km s$^{-1}$ were used to obtain the 
distance distribution and RV distribution, respectively. The red curves 
represents the best-fit Gaussian distributions. }\label{fig5}
\end{figure}

We excluded some member candidates with PMs larger than 
five times the standard deviation ($5\sigma$) from 
the mean PMs. This criterion allows us to select some 
walkaway stars. The mean and standard deviation values 
were redetermined using the remaining member candidates. 
This iterative process was performed until the statistical 
values reached constant values. 

Figure~\ref{fig3} displays the CMD of the selected 
members. However, the bright infrared source ($G_{\mathrm{RP}}$ = 7.1 
and $G_{\mathrm{BP}} - G_{\mathrm{RP}} = 3.2$) was still 
selected as a member. As we mentioned, this star may 
be an asymptotic giant branch star with similar 
kinematics to those of W5 members at almost the 
same distance. We excluded this star in our member list. 
A total of 490 candidates were finally selected as members.  
The members are listed in Table~\ref{tab1}.

Member candidates that we selected using optical and 
infrared data are active stars with warm circumstellar 
disks. There are, in fact, a number of diskless YSOs  in this 
SFR. Therefore, it is necessary to determine whether or not our 
member sample is representative for statistical analysis.

Such diskless YSOs without infrared excess emission 
have been identified using X-ray data in many previous studies 
\citep[etc]{GFB05,FMS06,TBC11,CMP12}. However, any extensive 
X-ray survey has not yet been performed for W5. Several 
hundreds of X-ray sources only in the eastern edge of W5 East 
(AFGL 4029) were detected \citep{TBG19}. We found a total 
of 257 X-ray counterparts in the Gaia EDR3 catalogue \citep{gedr3}, 
of which 56 were genuine members brighter than 18 mag 
in $G_{\mathrm{RP}}$ according to our member selection criteria. 
The member catalogue of this study contains 15 out of 56 X-ray 
sources. The PMs of members in the catalogue were 
compared with those of the 56 members with X-ray emission. 
As a result, they have median PM ($-0.110$ mas yr$^{-1}$, $-0.055$ 
mas yr$^{-1}$) similar to that of the X-ray members ($-0.122$ mas yr$^{-1}$, $-0.092$ 
mas yr$^{-1}$). Hence, we confirmed that the members selected in 
this study are representative sample of young stellar population in W5.

We present the spatial distribution of members in 
Figure~\ref{fig4}. W5 region has a high level of 
substructures. There are several groups of stars with 
high surface density and a distributed stellar population. 
The identification of stellar groups is addressed in a 
later section in detail. Figure~\ref{fig5} shows the 
distance and RV distributions of members. The distances 
of individual members were obtained from the inversion 
of the zero-point-corrected Gaia parallaxes \citep{gedr3,
LBB21}. These two distributions were fit to Gaussian 
distributions, respectively. We obtained the distance to W5 to be 
$2.1\pm 0.1$ (s.d.) kpc and its systemic RV to be 
$-37.8 \pm 3.3$ km s$^{-1}$ from the center values 
of the best-fit Gaussian distributions. RV data within 
three times the standard deviation from the mean RV 
were used to minimize the contribution of close binaries. 

\begin{deluxetable}{rccccccccccccccccc}
\rotate
\tabletypesize{\tiny}
\tablewidth{0pt}
\tablecaption{List of members \label{tab1}}
\tablehead{
\colhead{Sq.} & \colhead{R.A. (2000)}  & \colhead{decl. (2000)} &
\colhead{$\pi$} & \colhead{$\epsilon(\pi)$} & \colhead{$\mu_{\alpha}\cos \delta$} & \colhead{$\epsilon(\mu_{\alpha}\cos \delta)$} & \colhead{$\mu_{\delta}$} &
\colhead{$\epsilon(\mu_{\delta})$} & \colhead{RV$_{\mathrm{Helio}}$} & \colhead{RV$_{\mathrm{LSR}}$} & \colhead{$\epsilon(RV)$} &
\colhead{$G$} & \colhead{$G_{\mathrm{BP}}$} & \colhead{$G_{\mathrm{RP}}$} & \colhead{$G_{\mathrm{BP}} - G_{\mathrm{RP}}$} &
\colhead{Remark} & \colhead{Group} \\
        & \colhead{[deg]}  & \colhead{[deg]} &
\colhead{[mas]} & \colhead{[mas]} & \colhead{[mas yr$^{-1}$]} & \colhead{[mas yr$^{-1}$]} & \colhead{[mas yr$^{-1}$]} &
\colhead{[mas yr$^{-1}$]} & \colhead{[km s$^{-1}$]}& \colhead{[km s$^{-1}$]} & \colhead{[km s$^{-1}$]} &
\colhead{[mag]} & \colhead{[mag]} & \colhead{[mag]} & \colhead{[mag]} &
& }
\startdata
  1 & 41.146110 & 60.539714 & 0.2860 & 0.0905 &  0.445 & 0.088 & -0.605 & 0.087 & /nodata & /nodata &/nodata& 17.6966 & 18.8775 & 16.6265 & 2.2510 &     2 & A \\
  2 & 41.174792 & 60.702874 & 0.3841 & 0.1251 & -0.857 & 0.115 & -0.167 & 0.126 & /nodata & /nodata &/nodata& 18.0887 & 19.3473 & 16.9883 & 2.3590 &     2 & A \\
  3 & 41.175439 & 60.574290 & 0.6631 & 0.1922 & -0.277 & 0.192 & -0.514 & 0.188 & /nodata & /nodata &/nodata& 18.8903 & 20.2774 & 17.6886 & 2.5889 &     2 & A \\
  4 & 41.274583 & 60.566830 & 0.3946 & 0.0910 &  0.184 & 0.090 & -0.616 & 0.090 & /nodata & /nodata &/nodata& 17.6775 & 18.5560 & 16.7469 & 1.8091 &     2 & A \\
  5 & 41.313686 & 60.749972 & 0.2904 & 0.0934 & -0.414 & 0.088 &  0.299 & 0.087 & /nodata & /nodata &/nodata& 17.5657 & 18.5787 & 16.5543 & 2.0244 &     2 & A \\
  6 & 41.490458 & 60.703315 & 0.3637 & 0.1075 &  0.761 & 0.100 & -0.399 & 0.100 & /nodata & /nodata &/nodata& 17.8781 & 18.8807 & 16.9079 & 1.9729 &     2 & A \\
  7 & 41.577619 & 60.717774 & 0.6323 & 0.1784 & -0.067 & 0.153 & -0.279 & 0.151 & /nodata & /nodata &/nodata& 18.5599 & 19.7968 & 17.4335 & 2.3632 &     2 & A \\
  8 & 41.580421 & 60.793680 & 0.4743 & 0.0143 & -0.295 & 0.013 & -0.404 & 0.012 & /nodata & /nodata &/nodata& 13.3268 & 13.7419 & 12.7198 & 1.0220 &     T & A \\
  9 & 41.641083 & 60.662389 & 0.8520 & 0.0175 & -1.537 & 0.016 &  1.090 & 0.017 & /nodata & /nodata &/nodata& 12.0016 & 12.2946 & 11.5337 & 0.7609 & E     & A \\
 10 & 41.717295 & 60.223740 & 0.2982 & 0.0668 & -1.070 & 0.059 & -0.369 & 0.061 & /nodata & /nodata &/nodata& 17.0018 & 17.7457 & 16.1020 & 1.6437 &     2 & B \\
\enddata
\tablecomments{Column (1) : Sequential number. Columns (2) and (3) : The equatorial coordinates of members. Columns (4) and (5) : Absolute parallax and its standard error. Columns (6) and (7) : PM in the direction of right ascension and its standard error. Columns (8) and (9): PM in the direction of declination and its standard error. Columns (10 -- 12) : Heliocentric RV, RV in the local standard of rest frame, and its error. Column (13-15): $G$ magnitude, $G_{BP}$ magnitude, and $G_{RP}$ magnitude. Column (16) : $G_{BP} - G_{RP}$ color index. Column (17) : Classification of young stars. `E' represents O- or B-type stars obtained from the data bases of MK classification \citep[SIMBAD]{WOE00,R03,S09,MSM13}. `H' and `h' are H$\alpha$ emission stars and H$\alpha$ emission star candidates, respectively. `1', `2', and `T' denote Class I, Class II, and YSOs with a transitional disk, respectively. Column (18) : The host group name. The astrometric and photometric data were taken from the Gaia Early Data Release 3 \citep{gedr3}. This table is available in its entirety in machine-readable form.}
\end{deluxetable}

\begin{figure}[t]
\epsscale{1.0}
\plotone{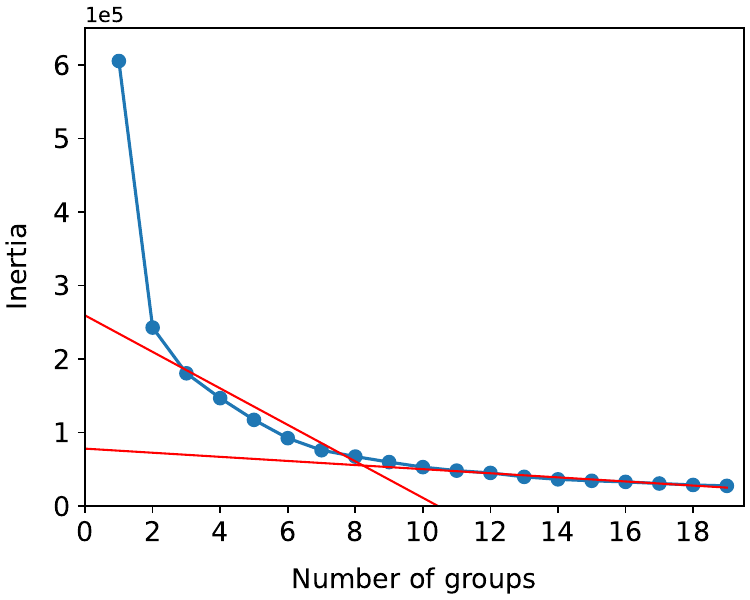}
\caption{Relation between the number of groups and inertia values. 
The number of groups was obtained from the elbow value of the 
inertia curve. The elbow value of eight (8) was determined from 
the point of intersection of the two red straight lines. See the main 
text for detail.}\label{fig6}
\end{figure}

\section{Results}\label{sec:4}

\subsection{Substructure}\label{ssec:41} 
Our previous studies have shown that stellar groups constituting 
the substructure in associations are spatially and kinematically 
distinct \citep{LNGR19,LHY20,LNH21,LNH22}. This means that they 
are individually different physical systems. We identified stellar 
groups in W5 by means of the unsupervised machine learning algorithm 
k-means clustering \citep{L82}. This algorithm finds a set of groups 
that have the smallest variance of each group (i.e., most compact groups) 
at a given number of groups. We used four-dimensional parameters as input data: R.A., decl., 
$\mu_{\alpha}\cos\delta$, and $\mu_{\delta}$. To find the optimized 
number of groups, we tested a number of groups from 1 to 20 and 
then computed the inertia value that is a sum of squared distances 
of stars to the centroid of their nearest group.

\begin{figure}[t]
\epsscale{1.0}
\plotone{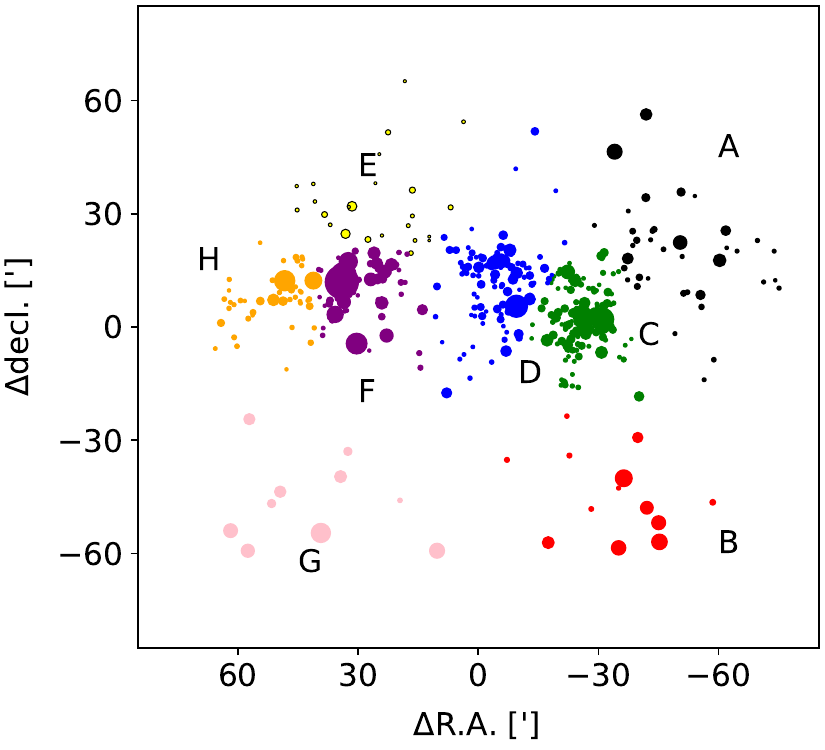}
\caption{Identification of stellar groups in W5. A total of 
eight groups were identified by means of the k-means clustering 
algorithm. The identified clusters were shown by different 
colors. The size of dots is proportional to the brightness 
of individual stars.}\label{fig7}
\end{figure}

Figure~\ref{fig6} displays the variation of the inertia values 
with respect to the number of stellar groups. Adopting a small number of 
groups prevents many real stellar groups from being identified, while 
adopting a large number of groups can lead to an overestimation of the 
number of genuine groups. In the figure, the location where an abrupt 
change of the inertia value occurs is referred to as the elbow. The elbow value 
is useful to determine the optimal number of groups. It is assumed that 
the variation of the inertia values approximates the combination of two 
straight lines (red lines in the figure). These two lines were obtained using a least square method 
for the number of groups ranging from 2 to 9 and 10 to 19, respectively.
The elbow value was determined as the intersection of two straight 
lines. Finally, we adopted eight groups that constitute the 
substructure in W5. The identified groups are plotted by different 
colors in Figure~\ref{fig7}. These stellar groups were named according 
to R.A. order (A to H). 

The groups C (green), D (blue), and F (purple) are stellar clusters 
denser than the other groups in W5. The two former 
groups are located at W5 West, while the latter ones are centered 
at the W5 East. Early-type stars in these three groups may 
be the main ionizing sources of W5. There are five sparse 
groups of stars around these clusters. The groups A, E, and 
H are located at the border of the northern H {\scriptsize II} 
bubbles, and the other two groups B and G are found in the 
southern part of the H {\scriptsize II} regions. We summarize 
the properties of individual groups in Table~\ref{tab2}.

In order to quantify the structural properties of the eight 
groups in W5, the minimum spanning tree (MST) technique were applied 
to the eight groups. The MST technique is to find the minimum length of 
the edges to connect all data points, which is often used to 
measure the degree of mass segregation \citep{AGP09}. In this 
work, we measured a dimensionless parameter $\Lambda$ defined 
as the ratio of standard deviation of length of edges to 
the mean length of edges \citep{HdGA17}. The larger $\Lambda$ 
means that there is a substructure in a group, such as core, 
clumps or filamentary structures. On the other hand, the lower 
$\Lambda$ means that there is no structural trend in the 
group and, for example, the $\Lambda$ becomes $\sim0.46$ when 
the group follows a uniform random distribution. The result of 
MST analysis can be found in Table~\ref{tab2}. Although 
the MST results of individual groups rely on the membership 
determination of host groups by the clustering algorithm, 
the results clearly show a trend that dense groups 
show larger $\Lambda$, and sparse groups show lower $\Lambda$. 
Especially, the MST results for three groups (B, E, and G) 
show that there is no significant structure.

We additionally tested three different clustering algorithms: 
Density-Based Spatial Clustering of Applications with Noise 
(DBSCAN; \citealt{EKS96}), hierarchical DBSCAN (HDBSCAN; \citealt{CMS13}), 
and Agglomerative Clustering. Adopting the results from the 
k-means clustering, we tuned parameters for each algorithm that 
makes similar results. Scikit-learn \citep{PVG11} was used for 
k-means clustering, DBSCAN, and Agglomerative Clustering, while 
the software developed by \cite{MHA17} was used for HDBSCAN.

DBSCAN finds groups based on the densities of data points. 
A main parameter is a radius ($\epsilon$) in consideration 
of neighboring data points. The $\epsilon$ value from 3.5 to 
3.0 identifies 7 to 10 groups. When $\epsilon$ is 3.5, 
DBSCAN identified two major groups in the east and west (D+C and F+H 
in Figure~\ref{fig7}). As $\epsilon$ decreases, these groups tend 
to be split. DBSCAN could not identify sparse groups (A, B, E, and G) 
in all cases; these group members were identified as noises.

DBSCAN is limited to identify groups with similar density, and 
therefore we implemented hierarchical HDBSCAN \citep{CMS13}. Unlike 
the DBSCAN algorithm, HDBSCAN can identify clusters with various density.
A major parameter is the minimum sample size ($n$), which is a number of 
neighboring points to be considered as cores. Larger $n$ provides 
more conservative clustering results. We tested $n$ from 5 to 20. 
When Similar to DBSCAN, HDBSCAN identified denser major groups in 
east and west (C, D, and H+F). These groups were split when using 
smaller $n$. Two sparse groups B and G were identified when 
using $n = 5$. Other sparse groups A and E were not identified in any cases.

We then tested a hierarchical clustering algorithm, 
Agglomerative Clustering. The major parameter is the number 
of groups. This algorithm was tested for the number of 
groups ($n_{\mathrm{groups}}$) from 4 to 14. When $n_{\mathrm{groups}}$=7, 
six groups (A, B, C, D, G, and H) were identified and 
the E+F group was identified as a single group. Larger $n_{\mathrm{groups}}$ 
values result in splitting groups such as A, D, and H. 
Smaller $n_{\mathrm{groups}}$ tends to combine groups; 
when $n_{\mathrm{groups}}$=4, group B and G are combined to C and F, respectively. 
While details are different, major 
results (east, west, and southern sparse groups) are similar 
to those obtained from the k-means clustering. 

All clustering algorithms that we used, in common, properly 
identified the dense groups although the memberships of stars 
at the boundaries of host groups is slightly different. The sparse 
groups were regarded as noise for the DBSCAN and HDBSCAN algorithms 
while the Agglomerative Clustering and k-means clustering 
algorithms identified them as real groups. Therefore, we 
should cautiously adopt the clustering results because the internal structures in SFRs are, 
in fact, very complex and related to their formation 
processes. Further information is required to 
determine if they are real physical systems. The ages and 
kinematics of group members may provide additional 
constraints on the identities of individual groups. 
Since some sparse groups seem to be real groups 
associated with remaining clouds, adopting the result 
from the k-means clustering is suitable for the purpose of this study.

\begin{table*}
\begin{center}
\setlength\tabcolsep{1.2pt}
\caption{Properties of individual groups \label{tab2}}
\begin{tabular}{ccccccccccccc}
\tableline\tableline\scriptsize
 Group & R.A. (2000) & decl. (2000)  & $\mu_{\alpha}\cos \delta_{\mathrm{med}}$ & $\mu_{\delta_{\mathrm{med}}}$  & $V_{\mathrm{R.A.,med}}$ & $V_{\mathrm{decl.,med}}$ &RV$_{\mathrm{LSR,med}}$ & $\sigma(V_{\mathrm{R.A.}})$ & $\sigma(V_{\mathrm{decl.}})$ & $\sigma$(RV$_{\mathrm{LSR}}$)& N & $\Lambda$\\
       & [deg]  & [deg]  & [mas yr$^{-1}$] & [mas yr$^{-1}$] & [km s$^{-1}$] & [km s$^{-1}$] & [km s$^{-1}$] & [km s$^{-1}$] & [km s$^{-1}$] & [km s$^{-1}$] &  & \\
\tableline
A & 41.996133 & 60.703094 & -0.215 &   -0.341 &  -2.1 & -3.4 & -29.1 & 5.3 & 2.8 & \nodata & 38(1) & 1.08 \\
B & 42.537271 & 59.593970 & -0.616 &   -0.479 &  -6.1 & -4.8 & \nodata & 6.9 & 6.1 & \nodata & 13(0) & 0.62 \\
C & 42.811804 & 60.393039 & -0.387 &   -0.442 &  -3.9 & -4.4 & -36.4 & 2.4 & 2.1 &   3.5   &159(15) & 1.13 \\
D & 43.522667 & 60.606917 & -0.138 &   -0.330 &  -1.4 & -3.3 & -38.3 & 3.5 & 2.4 &   2.7   &115(14) & 1.08 \\
E & 44.537468 & 60.884745 & -0.291 &   -0.044 &  -2.9 & -0.4 & -39.2 & 6.8 & 5.2 & \nodata & 25(4) & 0.87 \\
F & 44.785335 & 60.561629 & -0.277 &   -0.038 &  -2.8 & -0.4 & -39.0 & 2.4 & 2.0 &   2.4   & 84(25) & 1.12 \\
G & 45.152338 & 59.595502 &  0.205 &   -0.631 &   2.0 & -6.3 & \nodata & 5.6 & 8.7 & \nodata & 10(0) &0.74 \\
H & 45.331372 & 60.488579 & -0.110 &   -0.055 &  -1.1 & -0.5 & -37.4 & 3.1 & 3.8 &   0.6   & 46(12) & 1.73 \\
\tableline
\end{tabular}
\tablecomments{Column (1) : Group name. Columns (2) and (3) : Position of groups. Columns (4) and (5) : Median PMs along R.A. and decl. Columns (6--8) : Median tangential velocity along R.A., median tangential velocity along decl., and median RV. Columns (9--11) : Dispersion of tangential velocity along R.A., dispersion of tangential velocity along decl., and RV dispersion. Column (12) : The number of group members. The numbers in parenthesis represent the number of group members with RV measurements. Column (13) : Result of MST.}
\end{center}
\end{table*}

\begin{figure*}[t]
\epsscale{1.0}
\plottwo{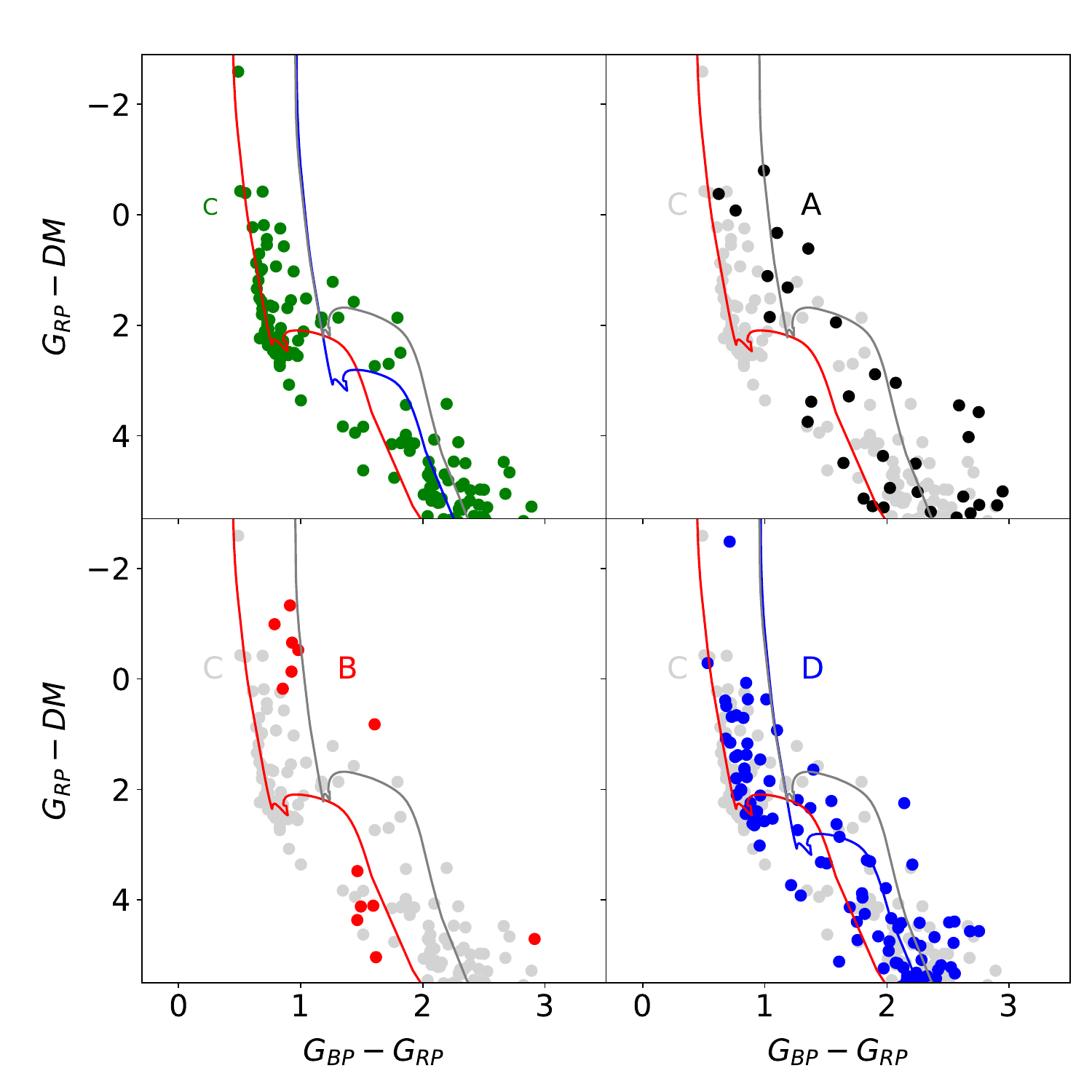}{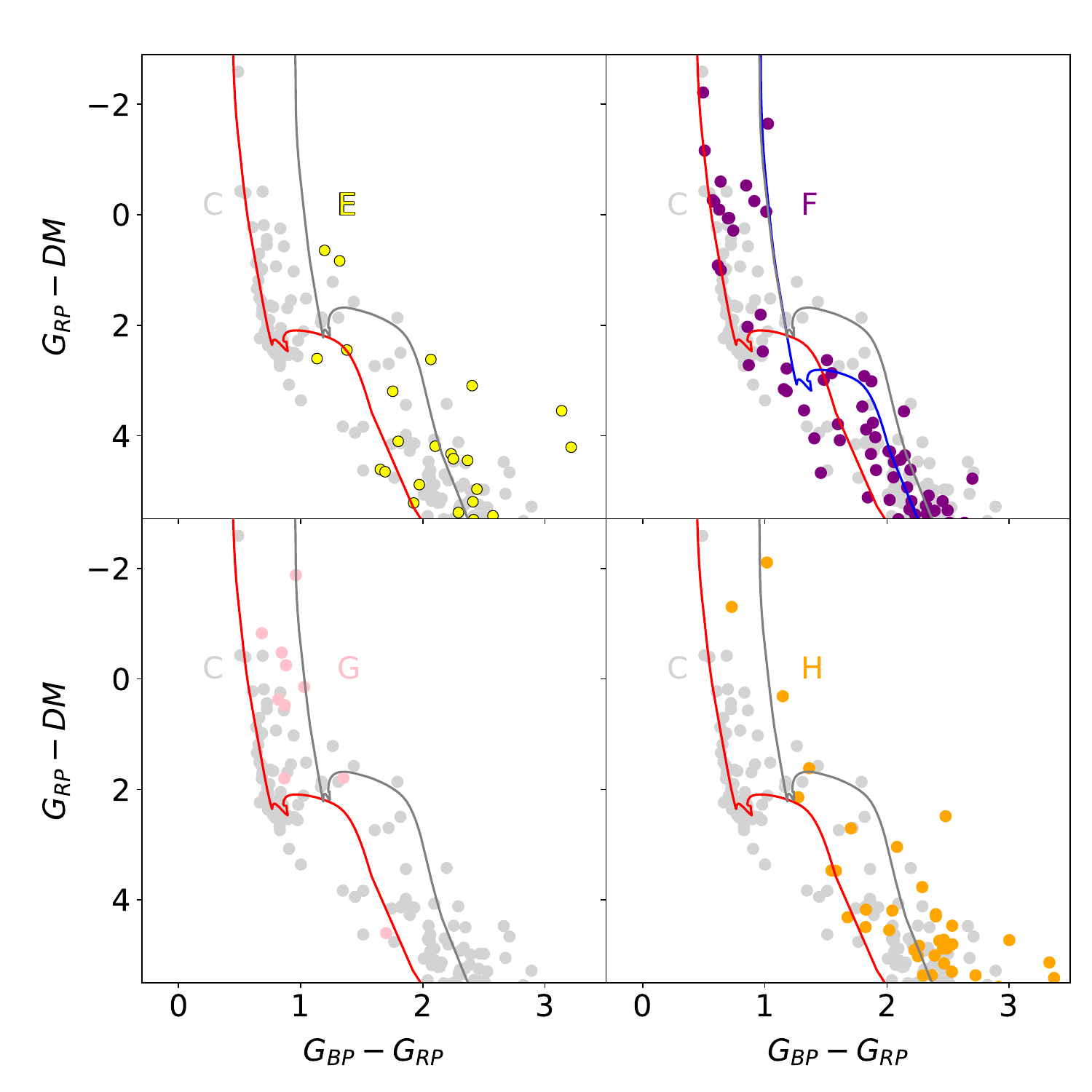}
\caption{CMD of stellar groups. The colors 
of dots represent the members of given groups corresponding 
to the color codes shown in Figure~\ref{fig7}. The CMD of the 
group C is plotted by gray dots for comparison. The red and blue 
curves exhibit the 5 Myr isochrones reddened by a total extinction 
($A_V$) of 1.86 and 3.00 mag, respectively, while the 2 Myr 
isochrone reddened by $A_V$ of 3.00 is plotted by gray curves. 
The CMDs of individual stellar groups are 
compared with that of the open cluster group C (green). }\label{fig8}
\end{figure*}

\subsection{Relative Ages}\label{ssec:42}
The star formation history in W5 can be inferred from the 
age distribution of stellar groups. The representative ages 
of individual groups are, in general, estimated from comparison 
of the overall features of CMDs with stellar evolutionary models. 
Especially, the luminosity of the main-sequence turn-on (MSTO) 
point (pre-main-sequence to main-sequence) is sensitive to the 
age of a given stellar group. 

Figure~\ref{fig8} displays the distance-corrected CMDs 
of individual groups with theoretical isochrones. The 
age of the group C (the open cluster IC 1848, green dots) 
is known to be about 1--5 Myr \citep{M72,KM03,KA11,
LSK14a}. We estimated the age of this cluster fitting the 
MESA isochrones considering the effects of stellar 
rotation \citep{CDC16,D16} to the CMD of the cluster. The 
distance modulus (DM) of 11.6 mag (2.1 kpc) was applied to 
the 5 Myr isochrone. The minimum extinction $A_V$ of 
1.86 mag was adopted from our previous study \citep{LSK14a}. 
This reddened isochrone fits well to the ridge of 
main-sequence as well as the luminosity of the MSTO. 
Therefore, we adopt 5 Myr as the age of this group. There are 
a number of stars brighter than the 5 Myr isochrone 
at given colors. These stars may be highly 
reddened stars (see the blue curve in the figure). 

We compared the MSTO and the overall features of the 
CMDs of individual stellar groups with that of the group 
C. The CMD morphology of the two dense groups D (blue) 
and F(purple) is very similar to that of the group C, 
and therefore the three dense groups would have similar age 
(5 Myr). The members of the sparse group A, E, and H are 
brighter and redder than those of the group C. We plotted 
the 2 Myr isochrones reddened by 3.00 mag, which 
passes through the middle of their CMDs. The color 
spread from the isochrones implies the presence of 
differential reddening across each group. In fact, 
they are located at the border of the H {\scriptsize II} 
region where large amounts of gas remain. This result 
indicates that there is an age difference between 
the northern sparse groups and dense groups. 
Therefore, the group division by the k-means clustering 
is meaningful.

On the other hand, the representative ages of the 
two southern groups B (red) and G (pink) are unclear as 
their MSTO is not well defined. The overall morphology 
of their CMDs are somewhat different from those of 
the other groups. There is no star between 2 and 4 mag in $G_{\mathrm{RP}} - DM$. 
The stars brighter than 2 mag close to the 2 Myr 
isochrone, while the faint stars seem 
to be older ($>$ 5 Myr) than the bright stars. We speculated 
that these two groups may not be the groups of coeval 
stars with the same origin.

\begin{figure}[t]
\epsscale{1.0}
\plotone{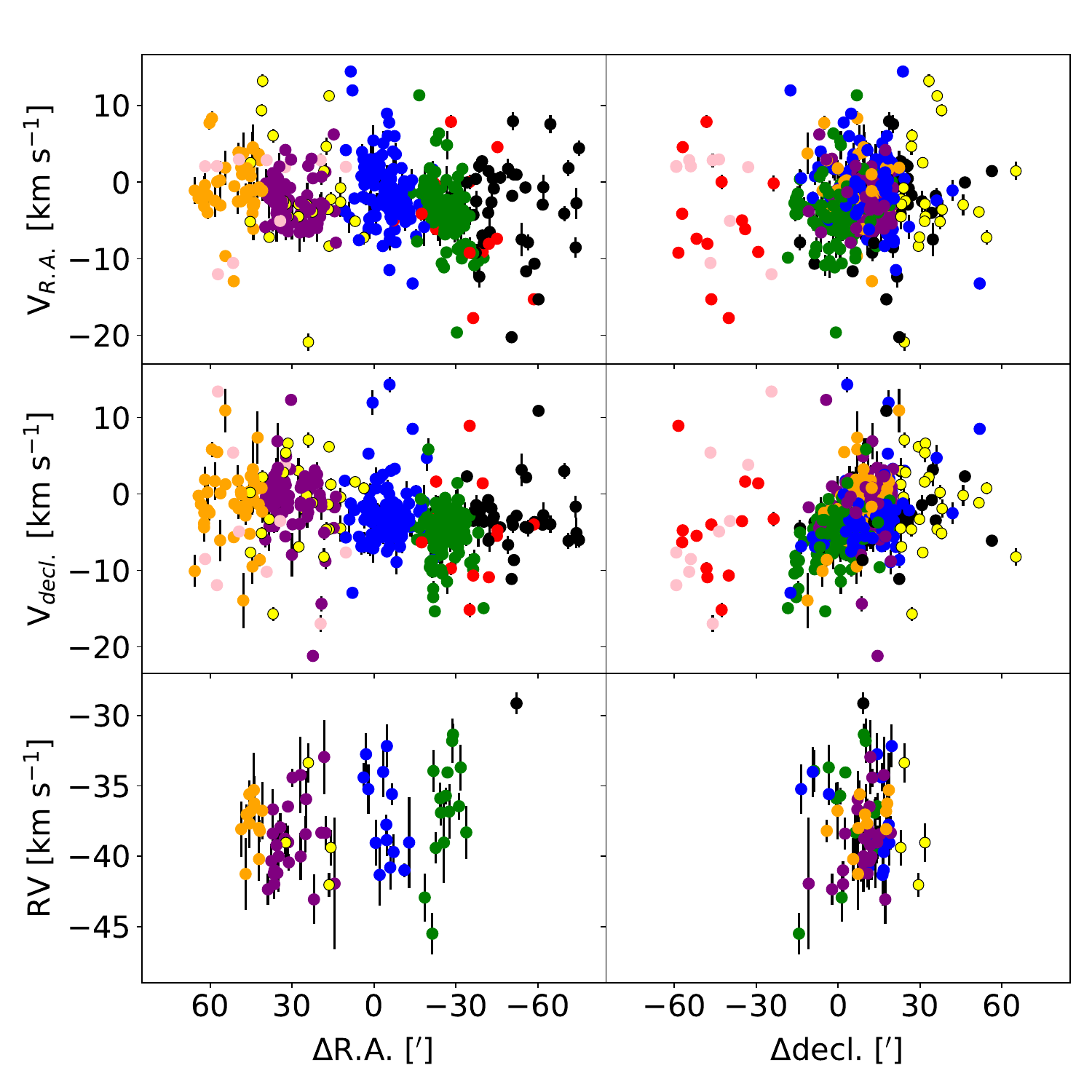}
\caption{Position-velocity diagrams of stars. The colors of 
dots are the same as those in Figure~\ref{fig7}. The vertical 
lines represent the errors of velocity measurements.}\label{fig9}
\end{figure}

\subsection{Kinematics}\label{ssec:43}
W5 has systemic PMs of $-0.273$ mas yr$^{-1}$ 
and $-0.333$ mas yr$^{-1}$ along R.A. and decl., 
respectively. We investigated the kinematics of 
individual groups. The tangential 
velocities ($V_{\mathrm{R.A.}}$ and $V_{\mathrm{decl.}}$) 
were computed from the PMs multiplied by the 
distance of 2.1 kpc. Figure~\ref{fig9} exhibits 
the distributions of velocities with respect 
to R.A. and decl. Since most groups are distributed along 
the east-west direction rather than the north-south 
direction, it is easier to probe some trends in 
position-velocity plane along R.A. 

There is no clear tendency between $V_{\mathrm{R.A.}}$ 
and positions of stars along R.A., while a gradual 
variation of $V_{\mathrm{decl.}}$ along R.A. is 
detected. $V_{\mathrm{decl.}}$ decrease at 0.08 
km s$^{-1}$ pc$^{-1}$. Any significant large-scale 
variation in RV was not found. We present the median 
velocities ($V_{\mathrm{R.A.,med}}$, $V_{\mathrm{decl.,med}}$, 
and RV$_{\mathrm{LSR,med}}$) in Table~\ref{tab2}. 

\begin{figure}[t]
\epsscale{1.0}
\plotone{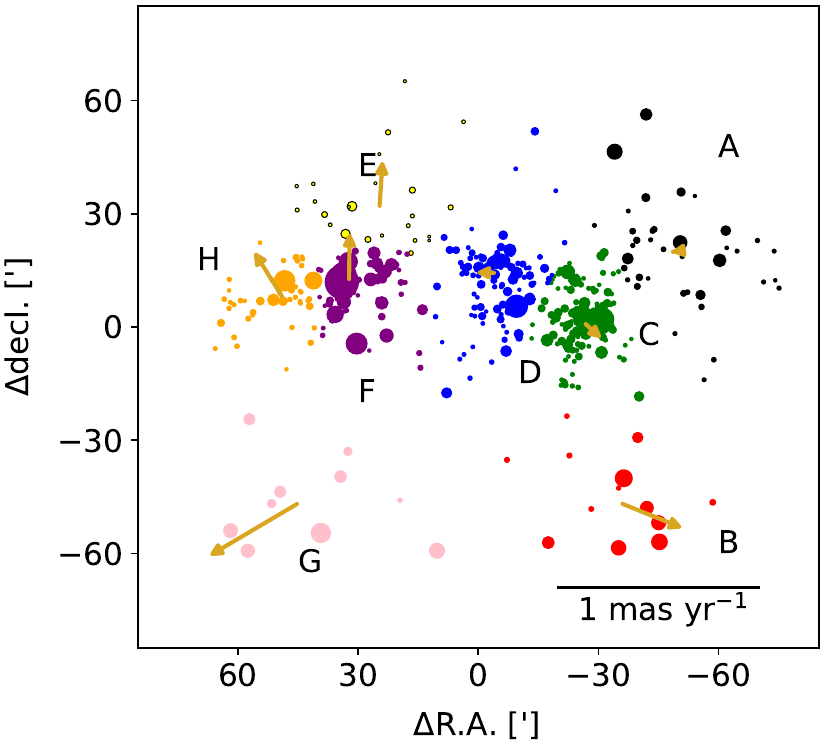}
\caption{Mean PMs of stellar groups relative to the systemic 
motion of W5. The brownish arrows represent the relative PM 
vectors of individual groups. The other symbols are the 
same as those of Figure~\ref{fig7}.}\label{fig10}
\end{figure}

We computed the standard deviation of the 
velocities of individual groups after excluding some 
outliers. The median measurement errors were adopted 
as the typical velocity errors. The velocity dispersions 
of given groups were then obtained from quadratic 
subtraction between the standard deviation values and 
the typical velocity errors. For RVs, we computed 
the velocity dispersions of the four groups that 
have more than ten stars with RV measurements. The 
velocity dispersions are presented in Table~\ref{tab2}.

The dense, populous groups C, D, and F tend to have 
velocity dispersions smaller than those of the other 
groups. In addition, the motions of stars in such groups 
seem to be nearly isotropic, given similar velocity 
dispersions among the tangential velocities and RV. 
The group A shows a large velocity dispersion in 
$V_{\mathrm{R.A.}}$, while it has a small value 
in $V_{\mathrm{decl.}}$ (see also Figure~\ref{fig9}). 
The groups B and G have particularly large velocity 
dispersions.

\begin{figure*}[t]
\includegraphics[width=16cm]{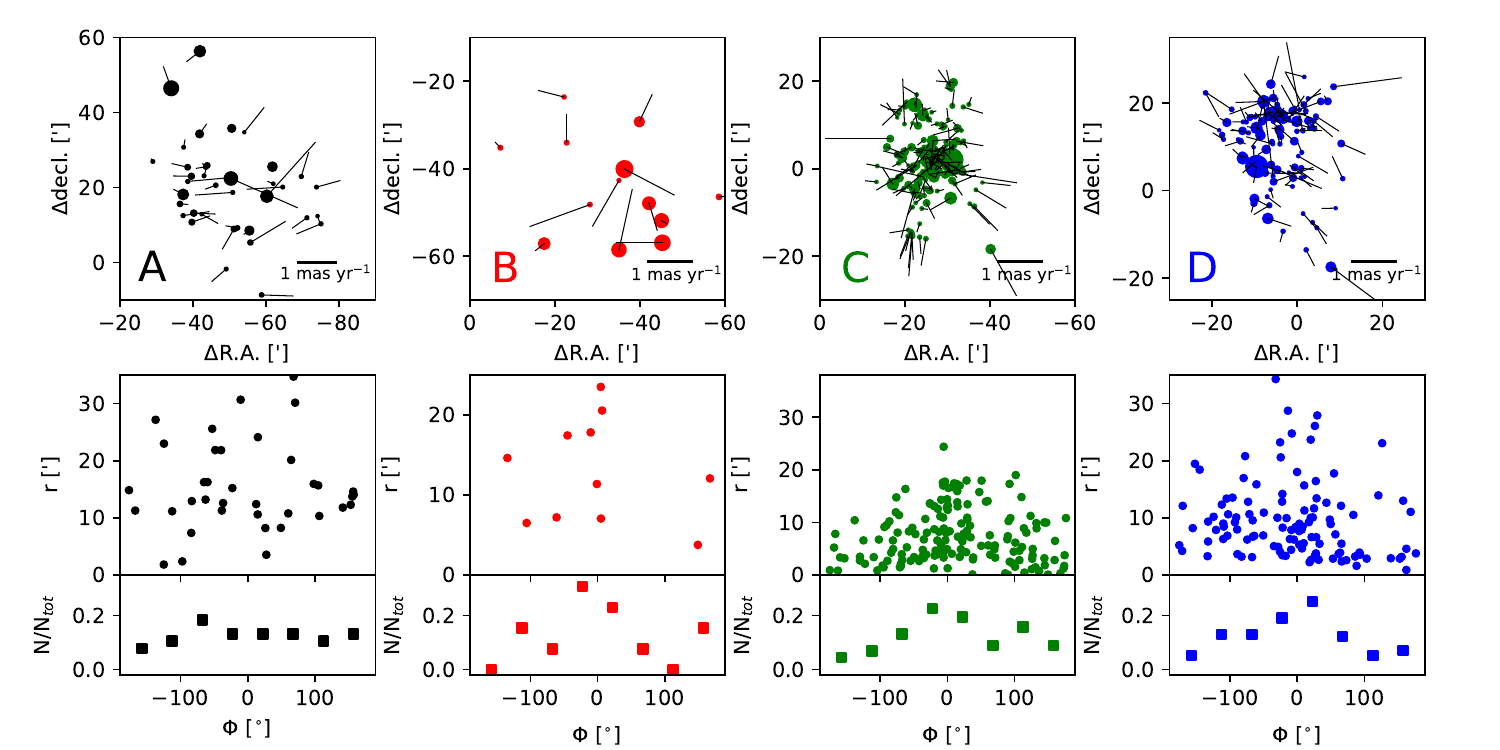}\\
\includegraphics[width=16cm]{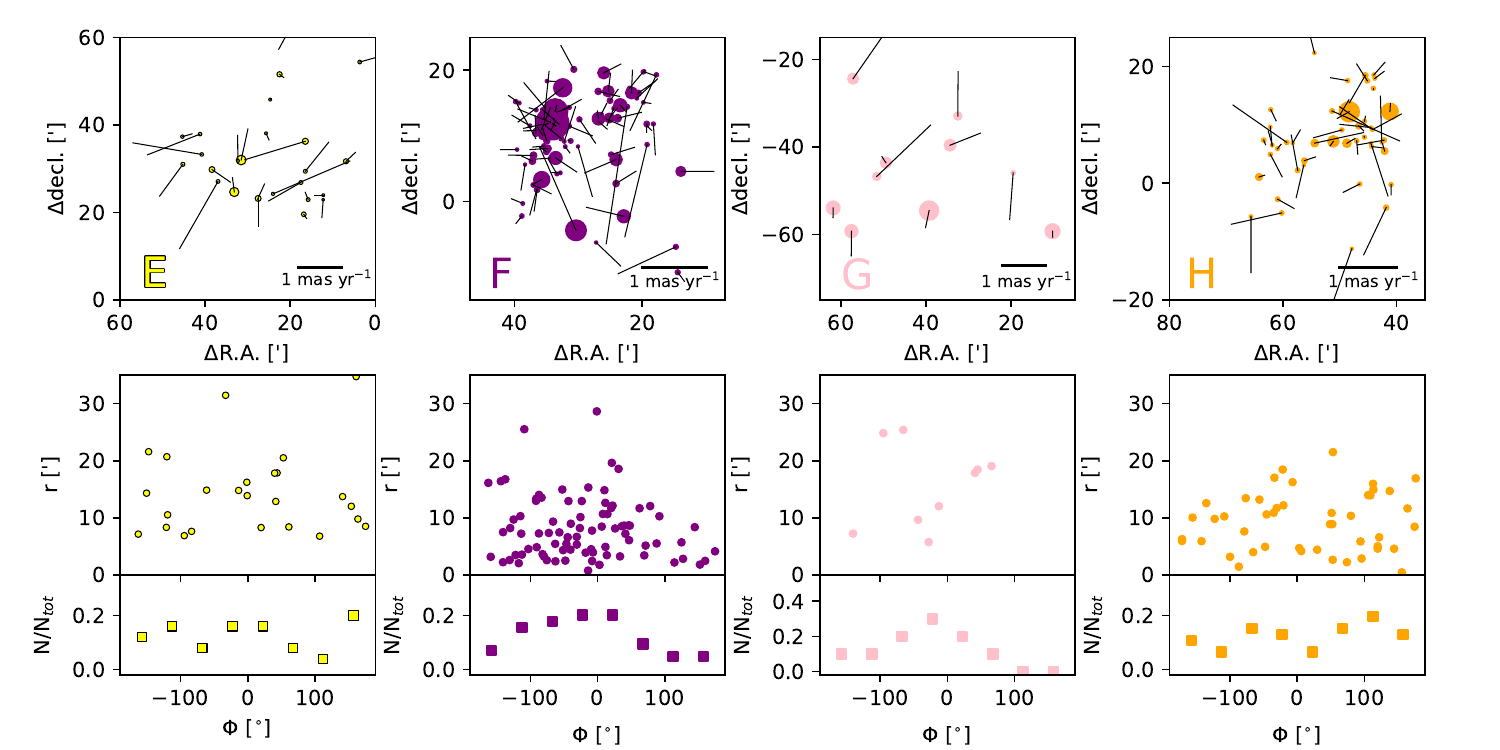}
\caption{Motions of stars in their host groups. The panels in 
the first and third rows display the spatial distributions 
of stars in given groups. The straight lines with different 
lengths represent the PM vectors of stars relative to their 
host groups. The $\Phi$ distributions are shown in the panels 
in the second and fourth rows. The colors of dots corresponds 
to those of stellar groups in Figure~\ref{fig7}.}\label{fig11}
\end{figure*}

We investigated the motions of individual groups 
within W5. Figure~\ref{fig10} shows the median PMs 
of individual groups relative to the systemic motion 
of W5. The number of members in W5 West is twice 
that in W5 East, so the groups A, 
C, and D have relative PMs close to the systemic 
motion of W5. The eastern groups E, F, and H are 
moving toward north, while the southern groups B 
and G are radially receding away from the center 
of this association.

Finally, the motion of individual members relative 
to the center of their host groups is probed in Figure~\ref{fig11}. 
The panels in the first and third rows of the figure 
exhibit the relative PM vectors of individual members 
in given groups. The group members show different 
pattern of motion. Some members of the groups C, D, 
and F seem to be moving outward from their group center. 
The PM vectors of members in the other groups have somewhat 
random direction. 

In order to quantitatively investigate the direction 
of their motion, we computed the vectorial angle ($\Phi$) 
that is defined by the angle between the position vector 
from the group center and the relative PM vector of 
members \citep{LNGR19,LHY20,LNH21,LNH22}. A $\Phi=0^{\circ}$ 
indicates that a star is radially receding away from 
the group center, while a $\Phi=180^{\circ}$ means that 
it is sinking inward. 

The panels in the second and fourth rows of 
Figure~\ref{fig11} displays the $\Phi$ distribution 
with respect to the projected distances from the center 
of each group. More than 40\% of members in the groups 
C, D, and F have $\Phi$ values close to $0^{\circ}$. 
This result indicates that these groups are expanding. 
The group B and G also shows a pattern of expansion, but 
the number of members are too small to confirm this 
claim. The members of the other groups show random 
$\Phi$ distributions, indicating random motion. The 
biggest difference between the dense groups and the two 
southern groups is that the dense groups have nearly 
isotropic inner region, which means that these groups 
are self gravitating.

\begin{figure*}[t]
\plottwo{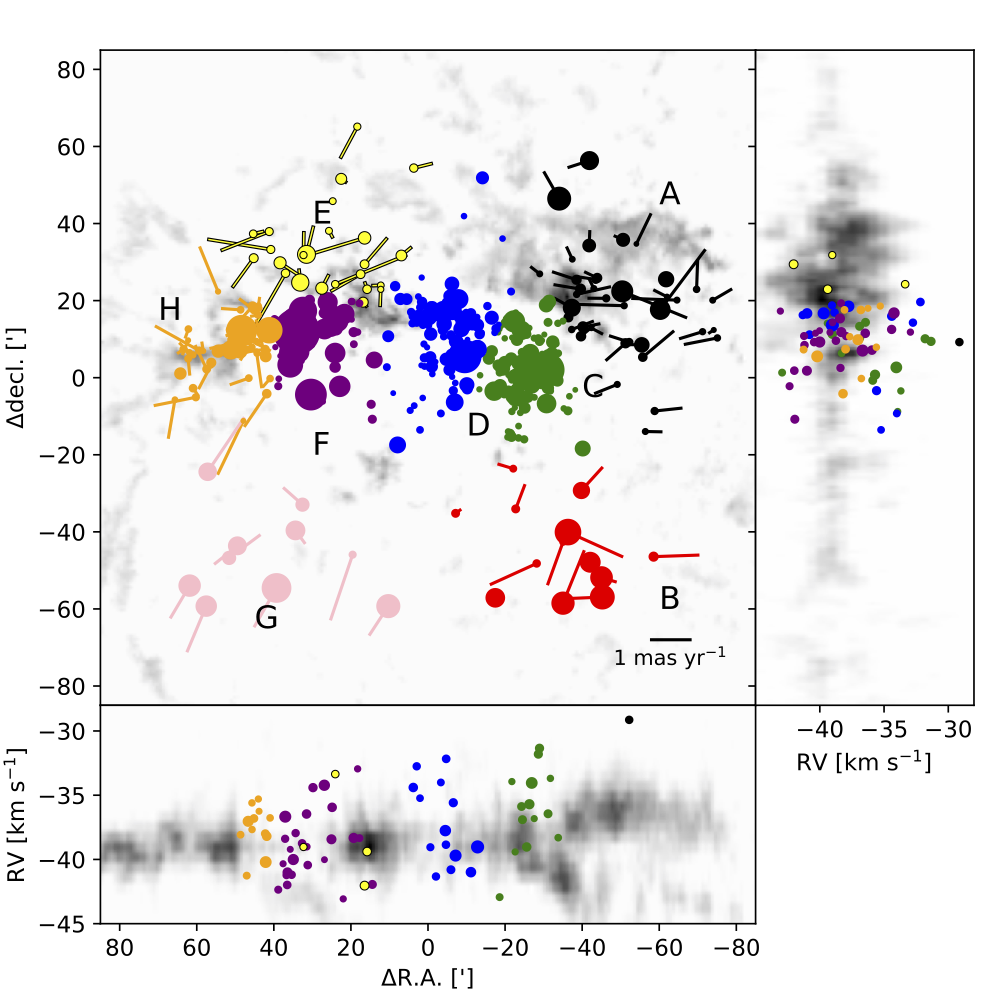}{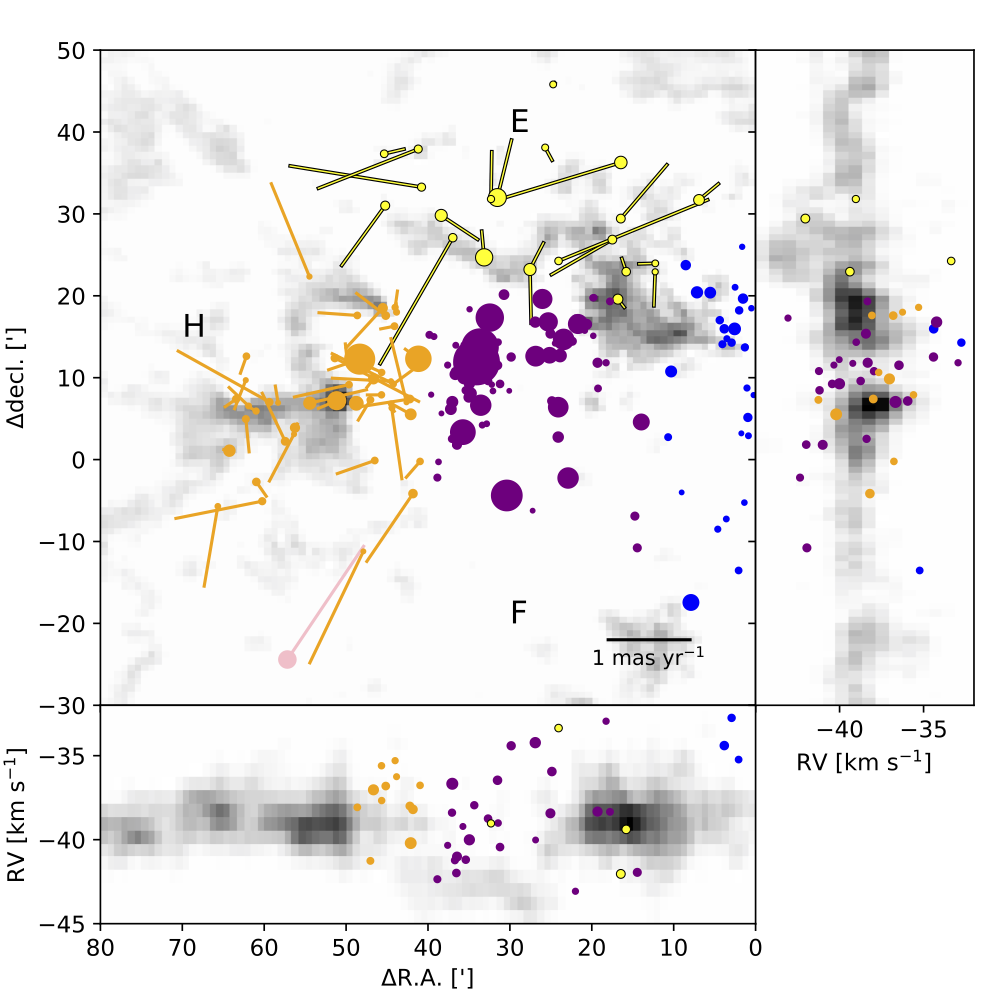}
\caption{Integrated intensity maps and PVs of the entire 
region (left) and W5 East region (right). These radio data were obtained 
from $^{12}$CO $(J = 1 - 0)$ line \citep{HBS98}. The 
gray scale represents the distribution of molecular 
clouds. The color-coded dots shows the distribution 
of members in the integrated intensity maps and PVs. 
The size of dots are proportional to the brightness 
of members. Arrows indicate the PM vectors of the 
sparse group members relative to their nearest dense 
groups (C in W5 West and F in W5 East). The nearest dense group containing ionizing sources 
in W5 West and East are the group C and F, respectively.}\label{fig12}
\end{figure*}

Figure~\ref{fig12} displays the integrated intensity 
map of the $^{12}$CO $J = 1 - 0$ taken from \citet{HBS98}. 
The integrated intensity map clearly shows the cavities in 
W5, where the groups C, D, and F incubating massive 
stars are located (see also \citealt{KAG08}). Most 
molecular clouds are distributed in the north of the 
three groups. The members of the stellar groups 
A, E, and H are spread over the northern clouds. 
Meanwhile, a small amount of clouds remain in 
the southern region.

The molecular clouds have RVs in a range of $-45$ 
km s$^{-1}$ to $-35$ km s$^{-1}$ as shown in the 
position-velocity (PV) diagrams of Figure~\ref{fig12}. 
Some clouds appear to be influenced by massive stars 
in groups C, D, and F. This aspect is found in 
the PV diagram along R.A. Although there is a scatter 
in stellar RVs, the RVs of molecular clouds are slightly 
smaller than those of the adjacent stellar groups 
hosting massive stars. Similar results were also 
found in some star-forming regions \citep{LSB18,
LNH21}.

\section{Star Formation in W5}\label{sec:5}
The extent of W5 is over 70 pc, and a high-level of 
substructures are found within the SFR. The dense 
stellar groups C, D, and F are located in the 
cavities of the giant H {\scriptsize II} regions. 
Their age estimated from the MSTO is, in common, about 
5 Myr, indicating that they formed in almost the same 
epochs. These dense groups are older than the other sparse 
groups, and therefore the star formation had been 
ignited at their current locations. 

These dense groups have velocity dispersions smaller 
than the other groups. If the small velocity dispersions 
indicate their physical state of their natal cloud, 
the cloud might have rapidly reached a subvirial state. Such a 
process may favorably occur in dense filaments by 
gravitational instability \citep{AMB10}, which may 
lead to cluster formation \citep{BSC11,K12}. Hence, 
the three dense groups might have formed in the densest region 
in a giant molecular cloud on a very short timescale.

About 40\% of stars in the dense groups are escaping 
from their host groups. Some stars scattered over 
this SFR may have originated from the expansion 
of such dense groups. However, their expansion pattern 
is not as significant as that found in IC 1805 which 
has a core-halo structure \citep{LHY20}. We therefore 
argue that the expansion of the dense groups may 
not be the origin of the overall structure in W5. 
The age difference between the dense groups and the 
northern groups supports this argument. Despite, it is 
expected that their expansion will lead to a distributed 
stellar population over several Myrs.

Star formation propagated to the northern part 
of W5 2 Myr ago. The sparse groups (A, E, and H) 
formed along the ridge of the H {\scriptsize II} 
regions. A number of previous studies have proposed 
that W5 is the site of feedback-driven star formation 
\citep{LW78,TTH80,WHL84,KM03,KAG08}. \citet{KAG08} 
suggested that the radiatively driven implosion mechanism \citep{KWS85} 
operates on a small spatial scale, e.g., cometary 
globules or elephant trunk structures in the southern 
ridge of W5 West, while the collect and collapse 
mechanism \citep{EL77} works on a larger scale, 
e.g., the northern clouds.

If the sparse groups were formed by the expansion 
of the H {\scriptsize II} regions, then 
they are expected to be receding away from ionizing 
sources. Since the group members form in compressed 
clouds, these stars have similar kinematics to that 
of the remaining clouds. However, observational 
results do not fully support the argument of 
\citet{KAG08}.

In figure~\ref{fig12}, the PM vectors of the group A 
members are shown relative to the group C. Their PM 
vectors have somewhat random orientation. Only one 
member has a RV measurement, and its RV is far different 
from that of the adjacent clouds. The RV data of 
more members would be required to better determine 
their systemic RV. 

We display the PM vectors of the group E and H members 
relative to the nearest dense group F in W5 East. 
Similarly, the group E members show random motions 
although they have similar RVs to those of the remaining 
clouds. The group H members are systemically receding 
away from the group F. We computed the $\Phi$ values 
of the group H members relative to the group H. As 
a result, the $\Phi$ distribution shows a peak at 
$\sim -20^{\circ}$. The fraction of stars showing such 
a systemic motion is over 50\% of all group members. 
Also, their RVs are consistent with those of 
the elephant trunks at the eastern edge of the 
H {\scriptsize II} region. The formation of the 
group H out of the three northern groups is likely 
associated with feedback from massive stars. 

Perhaps, the groups A and E spontaneously formed. 
They have velocity dispersions larger than those of 
the dense groups. It implies that their natal cloud 
was probably less favorable sites of star formation, 
like low-density thin filaments \citep{AMB10}. Star 
formation have proceeded on a timescale longer 
than the formation timescale of the dense groups.

Figure~\ref{fig12} also exhibits the PM vectors 
of the members in the southern groups B and G 
relative to the dense group C and F, respectively. 
The group B members have randomly oriented PM vectors, 
while more than 50\% of the group G members are 
moving away from this SFR. As seen in Section~\ref{ssec:42}, 
the group B and G members are not 
as young as those of the groups A, E, and H. Also, 
it is not sure that the members of these two groups 
are coeval populations as seen in their CMDs. 

Hence, these group members may have different 
origins. One possible explanation is that some 
of them are walkaway stars \citep{dMSL14}. Runaway 
and walkaway stars can originate from the end of 
binary evolution \citep{B61}. In this scenario, the massive primary 
star undergoes supernova explosion, and then the 
less massive secondary star is ejected, becoming 
either a runaway or a walkaway star. Another hypothesis 
is related to the dynamical ejection of 
stars in star-forming regions \citep{PRA67,OKP15,OK16}. 
Since there is weak evidence of supernova explosions 
in W5 \citep{VHV79}, the latter one is the 
more favorable mechanism for the southern groups in 
W5. The noncoeval population and the abnormal high-mass 
star population in CMDs (Figure~\ref{fig8}) can also be 
naturally explained according to this mechanism. 

\citet{LHY20} confirmed, in both numerical 
and observational ways, that subvirial collapse 
of cluster can lead to isotropic core and expanding 
halo structure. \citet{MPB22} also confirmed some 
stars and stellar systems isotropically ejected 
from the Bermuda cluster. They suggested that these 
ejected stars and stellar systems have a large amount of 
mass, and therefore such large mass-loss can result 
in cluster expansion. However, the group B and G members 
show anisotropic spatial distribution relative 
to this SFR. They occupy only the southern regions.  
If group G had the same origin as the halo of IC 1805 or 
the stellar systems ejected from the Bermuda cluster, 
it might miss some stars escaping in different 
directions, or the clustering algorithm we used 
might not be able to identify them. Based on 
current observational data, only a few stars 
are moving outward beyond this SFR.

The group B members have randomly-oriented PM vectors. 
The dynamical ejection mechanism cannot explain 
the random orientation in PM vectors. Low-levels of local star 
formation events may be another possible explanation 
for the origin of the southern groups. In this case, 
it should solve the question why there are more early-type 
stars than later-type stars, given the typical initial 
mass function \citep{Sa55,K01}.

In this study, a total of eight groups were 
identified. However, it is necessary to search for 
smaller subgroups in the future work. For instance, 
stars in the eastern part of group H ($\Delta$R.A. 
$\sim$ $60^{\prime}$) constitute a small aggregation. 
They seem to be associated with a small 
H {\scriptsize II} bubble east of the W5 East 
bubble (see figure 7 of \citealt{KAG08}). Their 
PM vectors relative to the group F show random 
directions, unlike stars in the western part of 
the group H. In addition, there is a small subgroup 
south of the group C. This group seems to be 
associated with the pillar-like structures 
at the border of the southern ridge of the W5 West 
bubble (see also figure 7 of \citealt{KAG08}). 
These further groupings will help to better 
understand the formation process of this SFR.

\section{Summary and Conclusion}\label{sec:6}
We studied the spatial and kinematic properties of 
young stars in the massive SFR W5 of the Cassiopeia 
OB6 association using the Gaia EDR3 data and 
high-resolution spectra to understand the formation 
process of stellar associations.

A total of 490 out of 2,000 young stars 
over W5 were selected as members using the Gaia 
parallaxes and PMs. The spatial distribution of 
the members reveals high-levels of substructures 
in W5. We identified eight stellar groups in total 
by means of the k-mean clustering algorithm. Three 
dense groups are centered at the cavities of the 
giant H {\scriptsize II} regions, and three sparse 
groups are found at the border of the H 
{\scriptsize II} bubble. The other two groups were 
found on the outskirt of the southern bubble. Our 
results were compared with those obtained from the 
other unsupervised machine learning algorithms, 
such as DBSCAN, HDBSCAN, and Agglomerative Clustering.

The dense groups are composed of the oldest stellar 
population (5 Myr), indicating that they are the first 
generation of stars in W5. They are now expanding. 
Three million years after their birth, star formation 
might have propagated toward 
the northern regions. Only one group (H) shows the 
signature of feedback-driven star formation. A number 
of its members are moving away from the nearest ionizing 
sources in the neighboring dense group F. In addition, 
their RVs are similar to those of the adjacent gas 
structures. On the other hand, the other two northern 
groups do not show such signatures, and therefore 
they might have spontaneously formed in the current positions.

The southern groups B and G seem not to be composed 
of coeval population. The group B members have 
randomly oriented PM vectors, while more than half 
of the group G members are moving away from W5. We 
discussed their possible origins as walkaway stars from W5 and/or 
multiple low-level star formation events. 

In conclusion, the major star formation process in W5 
may be associated with the structure formation in a 
giant molecular cloud. Multiple star formation 
might have spontaneously taken place in different 
positions and epochs. In addition, feedback from massive 
stars has triggered the formation of a new generation 
of stars, but the spatial scales at which this mechanism 
occurs may not as large as \citet{KAG08} suggested. 
Subsequent dynamical evolution of stellar 
groups will form a distributed stellar population in 
several Myr. 

\begin{acknowledgments}
The authors thank the anonymous referee for constructive 
comments and suggestions. The authors would also like to
express thanks to Prof. Mark Heyer for providing
supplementary data, Dr. Nelson Caldwell, and the other 
mountain staffs for assisting with Hectochelle observations.
Observations reported here were conducted at the MMT Observatory, 
a joint facility of the University of Arizona and the Smithsonian 
Institution. This paper has made use of data
obtained under the K-GMT Science Program (PIDs: MMT-2020B-001 and MMT-2021B-001) 
partly supported by the Korea Astronomy and Space Science
Institute (KASI) grant funded by the Korean government (MSIT;
No. 2023-1-860-02, International Optical Observatory Project)
and from the European Space Agency (ESA) mission Gaia
(https://www.cosmos.esa.int/gaia), processed by the Gaia Data
Processing and Analysis Consortium (DPAC, https://www.
cosmos.esa.int/web/gaia/dpac/consortium). Funding for the
DPAC has been provided by national institutions, in particular
the institutions participating in the Gaia Multilateral Agreement.
This research has also made use of the SIMBAD database,
operated at CDS, Strasbourg, France. This work was supported by
the National Research Foundation of Korea (NRF) grant
funded by the Korean government (MSIT; grant Nos. NRF2019R1C1C1005224
and 2022R1C1C2004102) and the research grant of Kongju National 
University in 2022. BL is grateful for Ms. Seulgi Kim’s 
assistance in data reduction and Prof. Jeong-Eun Lee's comments 
on observing proposal.
\end{acknowledgments}

\vspace{5mm}
\facilities{MMT:6.5m}


\software{{\tt xcsao} \citep{KM98}, {\tt NumPy} \citep{HMvdW20}, {\tt Scipy} \citep{VGO20}}








\end{document}